\documentclass[%
 reprint,
superscriptaddress,
 amsmath,amssymb,
 aps,
]{revtex4-2}

\usepackage{graphicx}
\usepackage{dcolumn}
\usepackage{bm}

\usepackage{adjustbox}
\usepackage{hyperref}
\usepackage{xcolor}
\usepackage{float}
\begin{document}

\title{Canonical spin glass behavior and memory effects in van der Waals magnet (Fe$_{0.55}$Ni$_{0.45}$)$_3$GaTe$_2$}

\author{P. Biswal}
\affiliation{Institute of Physics, Bhubaneswar 751005, India}
\affiliation{Homi Bhabha National Institute, Anushaktinagar, Mumbai 400094, India}
\author{Kakan Deb}
\affiliation{S. N. Bose National Centre for Basic Sciences, Salt Lake, Kolkata 700106, India}
\author{Nitesh Kumar}
\affiliation{S. N. Bose National Centre for Basic Sciences, Salt Lake, Kolkata 700106, India}
\author{D. Samal}
\email{Corresponding author: dsamal@iopb.res.in}
\affiliation{Institute of Physics, Bhubaneswar 751005, India}
\affiliation{Homi Bhabha National Institute, Anushaktinagar, Mumbai 400094, India}
\author{Prabhat Mandal}
\email{Corresponding author: prabhat.mandal@bose.res.in}
\affiliation{S. N. Bose National Centre for Basic Sciences, Salt Lake, Kolkata 700106, India}

\date{\today}

\begin{abstract}

We demonstrate the canonical spin-glass behavior and non-equilibrium dynamics in a single-crystalline Ni-doped van der Waals magnet (Fe$_{0.55}$Ni$_{0.45}$)$_3$GaTe$_2$ with a freezing temperature T$_f$$\simeq$45.5 K from detailed ac susceptibility and dc magnetization measurements. The temperature-dependent dc magnetization exhibits strong magnetic irreversibility and the ac susceptibility shows that the peak at T$_f$ shifts towards higher temperatures with increase in ac frequency, indicating the presence of a spin-glass state. The relative shift of the freezing temperature $\delta T_f$ per decade of frequency in ac susceptibility is estimated to be 0.0072. Dynamic scaling analysis of the frequency-dependent freezing temperature using the power law reveals the single spin flip relaxation time $\tau_0$=2$\times$10$^{-13}$ s and critical exponent z$\nu$=8.1, while the Vogel-Fulcher law yields Vogel-Fulcher temperature T$_0$=43 K and activation energy E$_a$/k$_B$=81 K with E$_a$/k$_B$T$_0$=1.88. The obtained values of these characteristic parameters lie well within the range associated with a canonical spin glass. The detailed non-equilibrium dynamics study, involving the magnetic relaxation and memory effect, further confirms the glassy behavior. The memory effect in zero-field-cooled magnetization establishes that the glassy state arises from the cooperative spin-spin interaction and not from superparamagnetism. The asymmetric response of magnetic relaxation with respect to negative and positive temperature cycle favors the hierarchical model of memory effects over the droplet model.

\end{abstract}

\maketitle


\section{Introduction}

Layered van der Waals magnets have drawn immense attention following the discovery of intrinsic two-dimensional (2D) ferromagnetism and layer-dependent ferromagnetism down to monolayer in Cr$_2$Ge$_2$Te$_6$~\cite{gong2017discovery} and CrI$_3$~\cite{huang2017layer}, respectively. Since then, several van der Waals magnets were discovered and these materials have emerged as promising platforms to explore exotic magnetic phenomena including noncollinear magnetism, topological magnetism, quantum anomalous Hall effect and provide excellent tunability in electronic, magnetic and structural properties using electrical, mechanical and chemical methods~\cite{yang2021van,liu2024magnetic,lian2025antiferromagnetic,deng2020quantum,kurebayashi2022magnetism,deng2018gate,kim2018large,seo2020nearly,mondal2021critical,wang2023interfacial,pal2024spin,pal2024disentangling,fei2018two}. Apart from long range magnetic order, van der Waals materials also exhibit spin glass state, originating from competing exchange interactions and quenched disorder present in the system~\cite{RevModPhys.58.801}. The defining characteristics of a spin glass phase include collective spin freezing, slow, non-equilibrium dynamics, history and frequency-dependent effects in the magnetic response, arising from the complex free energy landscape with multiple metastable states. The collective spin freezing can involve individual atomic spins (canonical spin glass) or correlated clusters of atomic moments (cluster spin glass)~\cite{kroder2019spin,PhysRevMaterials.7.034405,PhysRevB.98.144436}. Akin to conventional magnetic order, the stability of spin glass is intimately linked to the lattice dimensionality. For a 2D random Ising model, theoretical studies suggest that the  spin glass phase transition does not occur at any finite temperature and the correlation length diverges as temperature approaches to absolute zero~\cite{morgenstern1980magnetic}. For Ising systems with bimodal type disorder distribution, the lower critical dimension for a zero-temperature spin glass transition is two, whereas for Gaussian disorder it is above two~\cite{hartmann2001lower}. Moreover, for an XY and Heisenberg spin glass, the lower critical dimension for a spin glass phase transition lies above three~\cite{mcmillan1985domain,jain1986monte}. These theoretical results indicate that establishing a spin glass state becomes increasingly challenging in low-dimensional systems. However, from an experimental perspective, it is interesting to search for spin glass phenomenon in van der Waals materials with their dimensionality ranging from bulk single crystals to atomically thin layers.

\begin{figure*}
    \centering
    \includegraphics[width=0.8\linewidth]{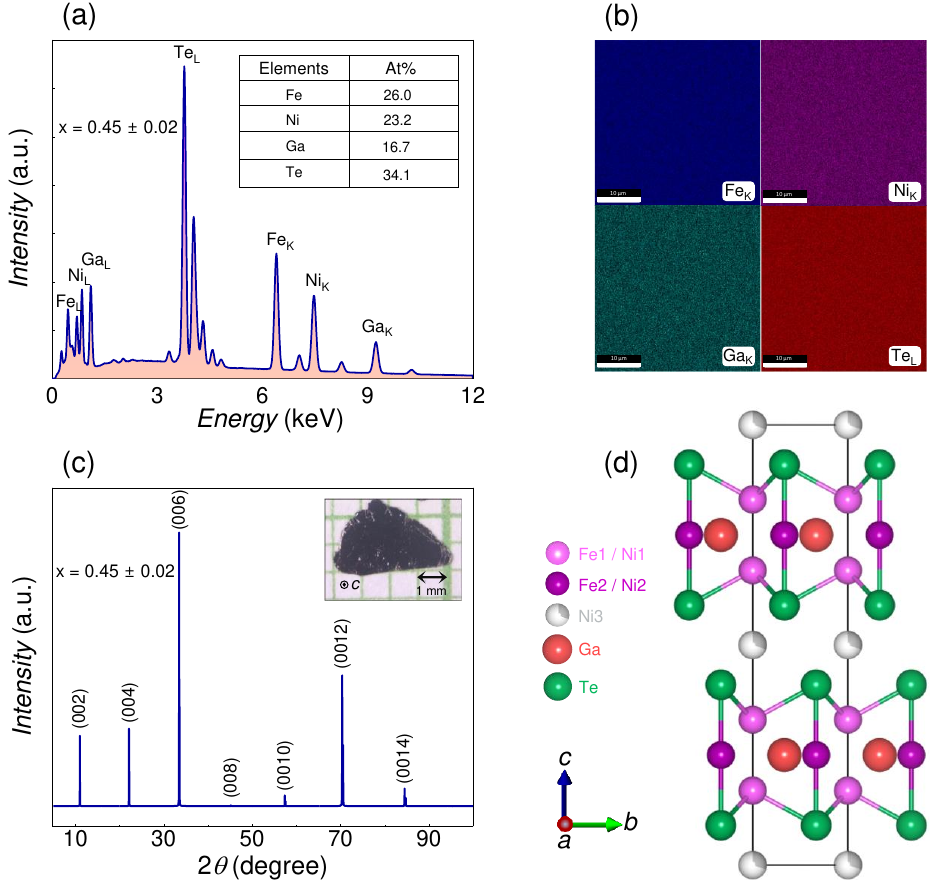}
    \caption{(a, b) EDXS spectrum (a) and elemental mapping images (b) of a Ni$_{0.45}$FGT single crystal. (c) Out-of-plane XRD pattern of Ni$_{0.45}$FGT. The inset shows the image of a Ni$_{0.45}$FGT single crystal. (d) Schematic representation of the crystal structure of Ni$_{0.45}$FGT viewed along the $a$ axis. }
    \label{structure}
\end{figure*}

Various approaches have been adopted to induce disorder and magnetic frustration to stabilize spin glass state in van der Waals compounds. Chemical doping has been used towards this goal as demonstrated in Mn$_{0.5}$Fe$_{0.5}$PS$_3$, Mn$_{1-x}$Fe$_x$PS$_3$, Fe$_{0.5}$Ni$_{0.5}$PS$_3$, Ru$_{1-x}$Cr$_x$Cl$_3$~\cite{takano2003magnetic,masubuchi2008phase,goossens2013magnetic,PhysRevB.99.214410}. Kurebayashi $\textit{et al.}$ found that Na intercalation in the van der Waals gap of Cr$_2$Ge$_2$Te$_6$ induces magnetic frustration, leading to a spin glass state~\cite{khan2024spin}. Incorporating structural disorder also offers a viable route. Zhang $\textit{et al.}$ observed spin glass behavior in amorphous Cr$_2$Ge$_2$Te$_6$, where amorphization leads to increased disorder arising from strong distortions of the Cr–Te–Cr bonds~\cite{wang2023spin}. Beyond externally induced disorder, there have been cases of observing an intrinsic spin glass state in van der Waals compounds. For instance, layered $\beta$-MoCl$_4$ exhibits spin glass state due to the highly disordered antiferromagnetic Mo lattice~\cite{mcguire2018spin}, whereas in Cr$_2$Se$_3$ nanosheets, glassy state arises owing to the spin frustration of partial Cr atoms~\cite{zhu2021spin}.

Interestingly Parkin \textit{et al.} in a recent study reported a spin-glass state in bulk Fe$_3$GeTe$_2$ compound, which persists to the 2D limit~\cite{pal2026realization}. The glassy state in Fe$_3$GeTe$_2$ is attributed to the dilute magnetic lattice formed by self-intercalated Fe atoms within the van der Waals gap. Prior to this study, spin glass state was observed in Fe$_3$GeTe$_2$ upon infusing a small percentage of randomly embedded FeTe clusters~\cite{BERA2023171052} and doping with Ni~\cite{drachuck2018effect}. Further, in the isostructural compound Fe$_3$GaTe$_2$, Chen \textit{et al.} reported a spin glass state upon Co and Ni doping~\cite{PhysRevB.109.104402}. The parent compound Fe$_3$GaTe$_2$ is a ferromagnet with a magnetic transition above room temperature and exhibits a robust perpendicular magnetic anisotropy~\cite{zhang2022above}. Substitution of Ni in Fe$_3$GaTe$_2$ transforms the ferromagnetic ground state to a spin glass state, possibly due to the structural disorder and magnetic dilution effects~\cite{PhysRevB.109.104402}. The glassy behavior in the doped (Ni$_x$Fe$_{1-x}$)$_3$GaTe$_2$ compound has so far been identified only from the frequency-dependent shift of ac susceptibility peak. However, a detailed investigation on the nature of the glassy state and related non-equilibrium dynamics in (Ni$_x$Fe$_{1-x}$)$_3$GaTe$_2$ remains unexplored.

In the present work, we establish the van der Waals magnet (Fe$_{0.55}$Ni$_{0.45}$)$_3$GaTe$_2$ as a canonical spin glass system  from comprehensive dc magnetization and ac susceptibility measurements. The relative shift in freezing temperature along with the characteristic parameters obtained from the power law and Vogel-Fulcher law analysis of the frequency-dependent peak shift in ac susceptibility, lie within the range defined for a canonical spin glass. The strong memory effect and magnetic relaxation studies using several experimental protocols further support the presence of glassy behavior. The memory effect observed in zero-field-cooled magnetization confirms that the glassy state arises from the cooperative spin-spin interaction and not from superparamagnetism. The asymmetric magnetic relaxation response with respect to negative and positive temperature cycle favors the hierarchical model of memory effects over the droplet model.

\section{Experimental Techniques and sample preparation}
Single crystals of $(\mathrm{Fe}_{0.55}\mathrm{Ni}_{0.45})_3\mathrm{GaTe}_2$ were grown by the self-flux method \cite{zhang2022above}. Fe powder (99.9\%), Ni powder (99.99\%), Ga lumps (99.9\%), and Te lumps (99.999\%) were used as starting materials in the molar ratio Fe:Ni:Ga:Te = 0.67:0.33:1:2. The mixture was loaded into an alumina crucible and subsequently sealed in an evacuated quartz tube. The sealed tube was heated to $1000^\circ\mathrm{C}$ at a rate of $200^\circ\mathrm{C/h}$ and maintained at this temperature for 24 h. The temperature was then reduced to $800^\circ\mathrm{C}$ at a rate of $100^\circ\mathrm{C/h}$, followed by slow cooling to $720^\circ\mathrm{C}$ with a rate of $5^\circ\mathrm{C/h}$ and held at $720^\circ\mathrm{C}$ for 50 h and finally cooled down to $350^\circ\mathrm{C}$ at a rate of $5^\circ\mathrm{C/h}$. Shiny plate-like single crystals with typical dimensions of $\approx 4 \times 3 \times 0.1~\mathrm{mm}^3$ were mechanically separated from the ingot. The actual chemical compositions were determined by energy-dispersive X-ray spectroscopy (EDXS, Quanta 250 FEG) measurements on a few randomly selected crystals, yielding an average ratio of Fe:Ni:Ga:Te = 26:23.2:16.7:34.1, consistent with the nominal stoichiometry of $(\mathrm{Fe}_{1-x}\mathrm{Ni}_{x})_3\mathrm{GaTe}_2$, $x = 0.45 \pm 0.02$ ($\mathrm{Ni_{0.45}FGT}$), as shown in FIG.~\ref{structure}(a). Moreover, the EDXS elemental mapping of $\mathrm{Ni_{0.45}FGT}$, presented in FIG.~\ref{structure}(b), indicates a homogeneous distribution of Fe, Ni, Ga and Te throughout the crystal. To further confirm the single-crystalline nature, x-ray diffraction (XRD) measurements were performed using a Rigaku SmartLab diffractometer with 9 kW Cu K$_\alpha$ radiation ($\lambda = 1.5418$ Å). Only sharp \((00l)\) diffraction peaks were observed, indicating that the crystals are oriented along the \(c\)-axis (FIG.~\ref{structure}(c)). Magnetic measurements were carried out on single crystals of $\mathrm{Ni_{0.45}FGT}$ using a superconducting quantum interference device based magnetometer (Quantum Design SQUID‑VSM).

\section{Results and discussion}

\subsection{Crystal structure}
Fe$_3$GaTe$_2$ crystallizes in a hexagonal structure with the space group $P6_3/mmc$ (No. 194). Two crystallographically inequivalent Fe sites, denoted as Fe1 and Fe2, are present in the structure. The unit cell consists of two Fe$_3$GaTe$_2$ slabs stacked along the $c$ axis and separated by a van der Waals gap. Within each slab, the Fe2-Ga layer is sandwiched between Fe1 and Te layers \cite{zhang2022above}. Previous studies on Ni-doped Fe$_3$GaTe$_2$ have shown that Ni atoms initially occupy the Ni3 site located within the van der Waals gap, which is only partially occupied, and with increasing Ni concentration, Ni atoms preferentially occupy the Fe2 site. Ni starts to occupy the Fe1 site at higher doping levels \cite{yuan2025direct}. Therefore, for Ni$_{0.45}$FGT, the Ni atoms are likely to be distributed over the Ni3 and Fe2 sites, with a possible contribution from the Fe1 site (FIG.~\ref{structure}(d)).
\begin{figure}
    \centering
    \includegraphics[scale=0.37]{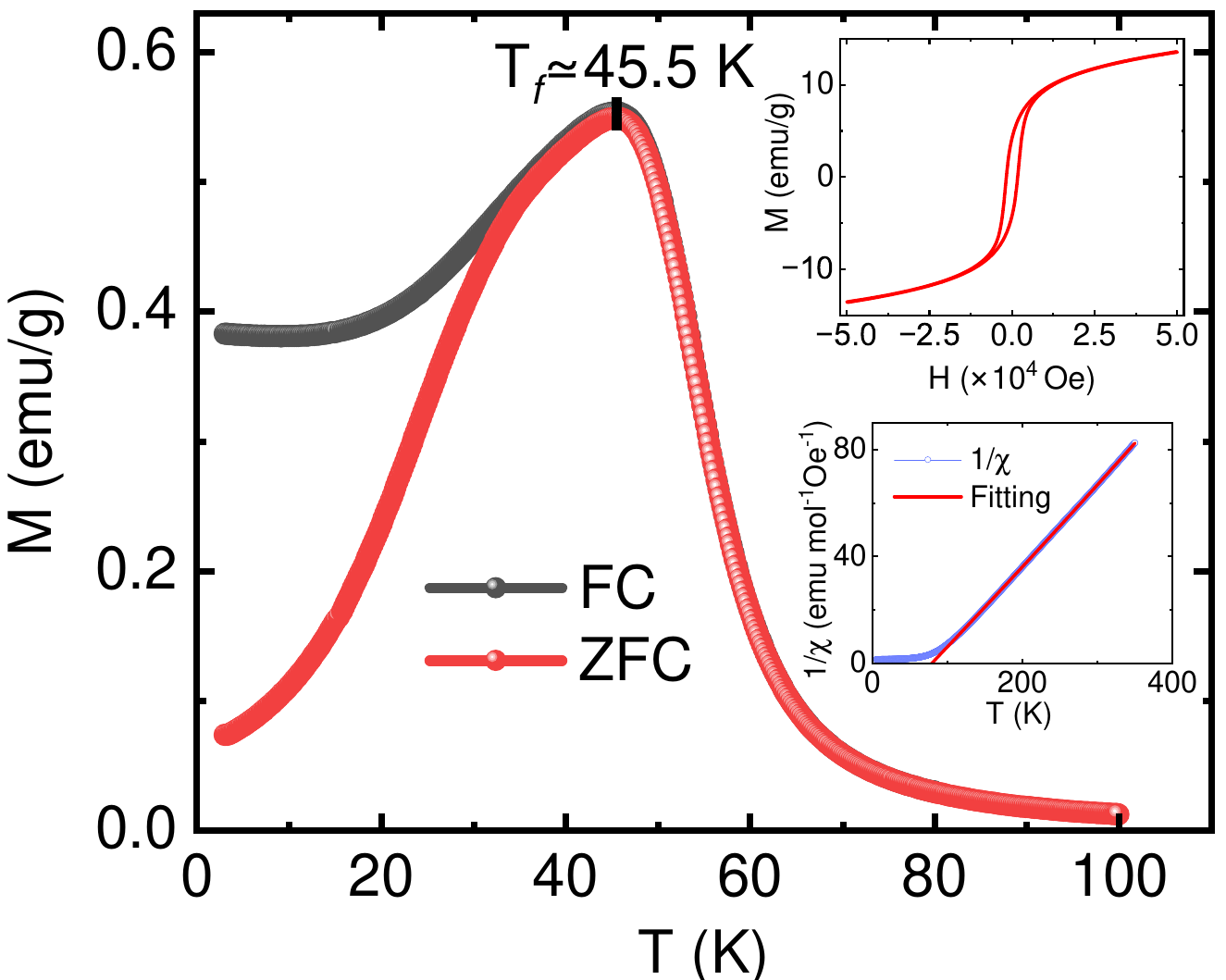}
   \caption{Temperature-dependent magnetization (M (T)) measured in field cooling (FC) and zero-field cooling (ZFC) mode under an applied field of 50 Oe. The upper inset shows field-dependent magnetization M(H) measured at T=5 K. The lower inset shows the inverse of susceptibility (1/$\chi$(T)) at an applied field of 1 Tesla measured in field cooling (FC) mode. The red line represents the fitting following the modified Curie-Weiss law $\chi=\chi_0+\frac{C}{T-\theta_{CW}}$. }
    \label{Figure chi}
\end{figure}


\subsection{dc magnetization}

FIG.~\ref{Figure chi} shows the temperature-dependent dc magnetization M, data collected in zero-field cooling (ZFC) and field cooling (FC) mode under an applied field (H) of 50 Oe. Both curves show a maximum around 45.5 K, below which they show a downfall and bifurcate from each other. Further, a field-dependent magnetization M(H) at T=5 K (the upper inset of FIG.~\ref{Figure chi}) exhibits a small hysteresis with no indication of saturation up to 5 Tesla. The strong magnetic irreversibility manifested through ZFC-FC bifurcation and the non-saturating magnetic hysteresis, indicates the possible existence of a spin-glass-like phase in Ni$_{0.45}$FGT with a spin-freezing temperature T$_f$$\simeq$45.5 K~\cite{gupta2023experimental}. In the subsequent sections, a comprehensive investigation employing frequency-dependent ac susceptibility along with memory and aging effects has been undertaken to firmly establish the spin-glass ground state of Ni$_{0.45}$FGT.

The nature of the underlying dominant magnetic interaction in this system can be determined from the value of Curie-Weiss temperature~\cite{mugiraneza2022tutorial}. The lower inset of FIG.~\ref{Figure chi} shows the temperature-dependent inverse susceptibility (1/$\chi$(T)) at an applied field of 1 Tesla, collected in FC mode. Fitting of the 1/$\chi$(T) with the inverse of modified Curie-Weiss law $\chi=\chi_0+\frac{C}{T-\theta_{CW}}$ (where $\chi_0$ is the temperature independent susceptibility, C is the Curie constant and $\theta_{CW}$ is the Curie-Weiss temperature)~\cite{mugiraneza2022tutorial} yields $\chi_0$=-4.068$\times$10$^{-4}$ emu mol$^{-1}$Oe$^{-1}$, C=3.402(3) emu mol$^{-1}$Oe$^{-1}$ K and $\theta_{CW}$=78.60(8) K. The positive value of $\theta_{CW}$ suggests the presence of ferromagnetic interaction in the van der Waals compound Ni$_{0.45}$FGT.

\begin{figure}
    \centering
    \includegraphics[scale=0.34]{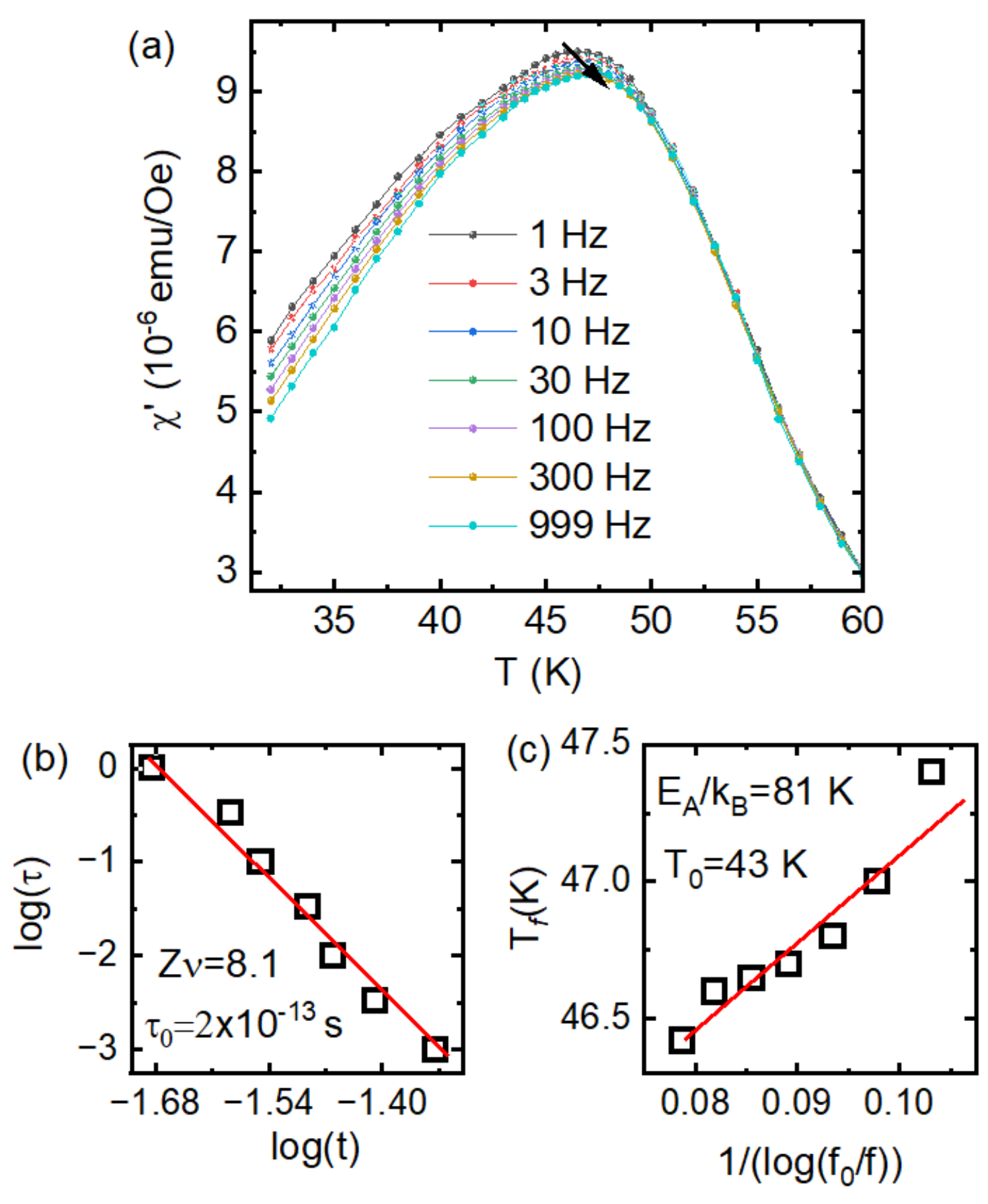}
   \caption{(a) Temperature-dependent real part of ac magnetic susceptibility in the frequency range of 1 Hz to 999 Hz. (b) The reduced temperature (t) dependent relaxation time ($\tau$) on a log-log scale. The red line represents the linear fit. (c) Frequency-dependent freezing temperature (T$_{f}$) according to the Vogel-Fulcher law. The red line represents the linear fit.}
    \label{ac chi}
\end{figure}

\subsection{ac magnetic susceptibility}

To examine the possible existence of a glassy spin state in Ni$_{0.45}$FGT, temperature-dependent ac magnetic susceptibility is measured in the frequency ($f$) range of 1-999 Hz at an excitation field (H$_{ac}$) of 2 Oe. In FIG.~\ref{ac chi}(a), the real part of ac susceptibility ($\chi'$(T)) shows a peak similar to that observed in the temperature-dependent low-field dc magnetization curves (FIG.~\ref{Figure chi}). The peak in $\chi'$(T), gradually shifts to higher temperature side with increase in frequency, i.e., T$_{f}$ increases with $f$. Such a frequency-dependent peak shift indicates the presence of a spin-glass state in Ni$_{0.45}$FGT.

The relative shift of T$_{f}$ per decade of frequency in a spin glass state is quantified using the Mydosh parameter ($\delta T_f$) ~\cite{PhysRevB.25.515,mydosh1993spin}, which is calculated as follows; 
\begin{equation}
\delta T_f=\frac{\Delta T_f}{T_f\Delta(log_{10}f)}
    \label{eq myodosh parameter}
\end{equation}
where, $\Delta T_f$=(T$_f$)$_{f_1}$-(T$_f$)$_{f_2}$ and $\Delta(log_{10}f)$=(log$_{10}f_1$-log$_{10}f_2$). By considering $f_1$=1 Hz and $f_2$=999 Hz, the Mydosh parameter $\delta T_f$ is found to be 0.0072, which is comparable to the values reported for canonical spin glasses such as AuMn ($\delta T_f$=0.0045), CuMn ($\delta T_f$=0.005), AgMn ($\delta T_f$=0.006) and AuFe ($\delta T_f$=0.01)~\cite{PhysRevB.25.515,tien20002,mydosh1993spin}.


In a spin glass system, the standard power-law divergence associated with a critical slowing down of spin dynamics is given as~\cite{hohenberg1977theory,mydosh1993spin}

\begin{equation}
\tau=\tau_0(\frac{T_f-T_{SG}}{T_{SG}})^{-z\nu}=\tau_0t^{-z\nu}
    \label{eq 1}
\end{equation}

where $\tau$ is the relaxation time associated with the corresponding measured frequency ($\tau=1/f$), $\tau_0$ is the characteristic relaxation time for a single spin flip, T$_{SG}$ is the spin glass freezing temperature at f=0, $t$ [=$(T_f-T_{SG})/T_{SG})$] is the reduced temperature, and z$\nu$ is the dynamic critical exponent [$\nu$ is the critical exponent of correlation length, $\zeta=t^{-\nu}$ and $\tau\sim\zeta^z$]. The FIG.~\ref{ac chi}(b) shows the $\tau$ vs. $t$  in the log-log scale. From the linear fitting using equation~\ref{eq 1}, the $\tau_0$ and z$\nu$ are obtained to be 2$\times$10$^{-13}$ s and 8.1 respectively. For spin glass systems, generally z$\nu$ lies between 4-12, while the $\tau_0$ is of the order of 10$^{-12}$-10$^{-13}$ s for canonical spin glass and 10$^{-7}$-10$^{-10}$ s for cluster glass systems~\cite{laiho2001spin,malinowski2011spin,nam2000spin,ghara2014reentrant,PhysRevB.98.144436}. The estimated values of $\tau_0$ and z$\nu$ for Ni$_{0.45}$FGT, lie within the range expected for canonical spin glass systems.

Vogel-Fulcher law provides another description for spin glass freezing dynamics by taking spin interactions into account~\cite{PhysRevB.32.516} and it is described as   
\begin{equation}
f=f_0 exp(\frac{-E_a}{k_B(T_f-T_0)})
    \label{eq 2}
\end{equation}
where $f_0$ is the characteristic frequency, $E_a$ is the activation energy and T$_0$ is the Vogel-Fulcher temperature. Considering equation~\ref{eq 2}, the frequency dependence of T$_{f}$ can be presented as 
\begin{equation}
T_f=T_0+\frac{(E_A/k_B)}{ln(f_0/f)}
    \label{eq 3}
\end{equation}
According to the above equation~\ref{eq 3}, the linear fitting shown in FIG.~\ref{ac chi}(c) yields $E_a/k_B$= 81 K and T$_0$= 43 K. Tholence introduced the parameter $\alpha=(T_f-T_0)/T_f$ for spin glass systems. Here, $\alpha$ is obtained to be 0.055, which lies well within the expected range for spin glass systems \cite{yeshurun1985spin,tholence1984recent}. The ratio of activation energy to T$_0$ determines the strength of interaction between dynamical entities freezing at T$_f$. Here, $E_a/k_BT_0$ is estimated to be 1.88, which again lies around the range 2-3, expected for canonical spin glasses~\cite{khan2013critical,lago2012three}. These results suggest that Ni$_{0.45}$FGT is a canonical spin glass system.

\begin{figure}
    \centering
    \includegraphics[scale=0.36]{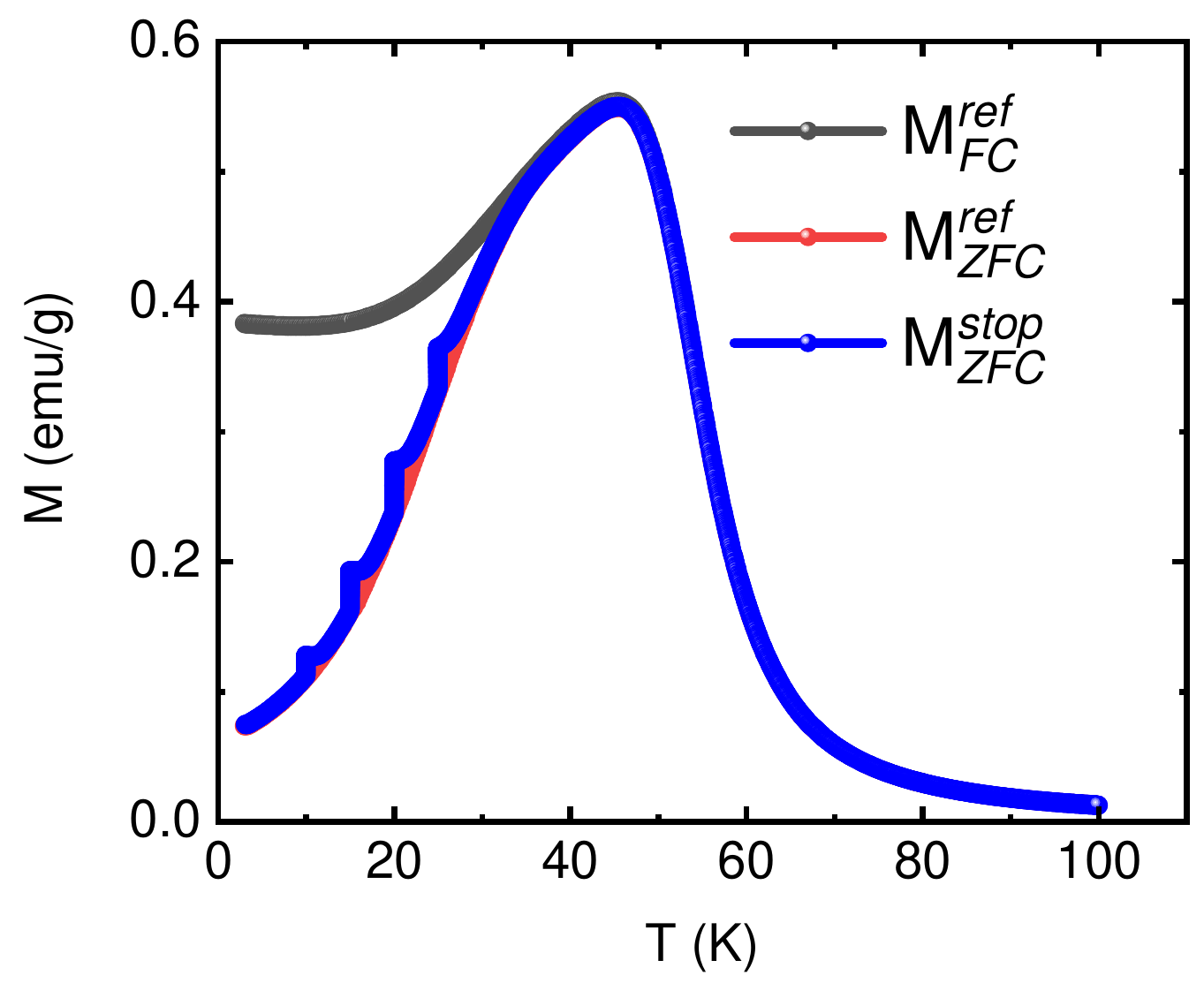}
   \caption{Temperature dependent field cooled (M$_{FC}^{ref}$) and zero field cooled (M$_{ZFC}^{ref}$) dc magnetization at 50 Oe. The step like zero field cooled magnetization M$_{ZFC}^{stop}$ is due to the intermediate halts with a duration of 1 h each at temperatures of 10, 15, 20, and 25 K.}
    \label{Figure 1}
\end{figure}

\subsection{Nonequillibrium dynamics}

\subsubsection{\textbf{Halting-dependent step-like features in magnetization}}
 
Temperature-dependent field-cooled (M$_{FC}^{ref}$) and zero field cooled (M$_{ZFC}^{ref}$) magnetization collected in the warming mode for an applied field of 50 Oe are shown in FIG.~\ref{Figure 1} along with another zero-field-cooled magnetization (M$_{ZFC}^{stop}$) recorded while halting at each temperatures 10, 15, 20, 25 K for 1 h. Prominent step-like features at these halting temperatures are observed in M$_{ZFC}^{stop}$, showing the time evolution and glassy dynamics of this system below T$_f$.


\begin{figure}
    \centering
    \includegraphics[scale=0.35]{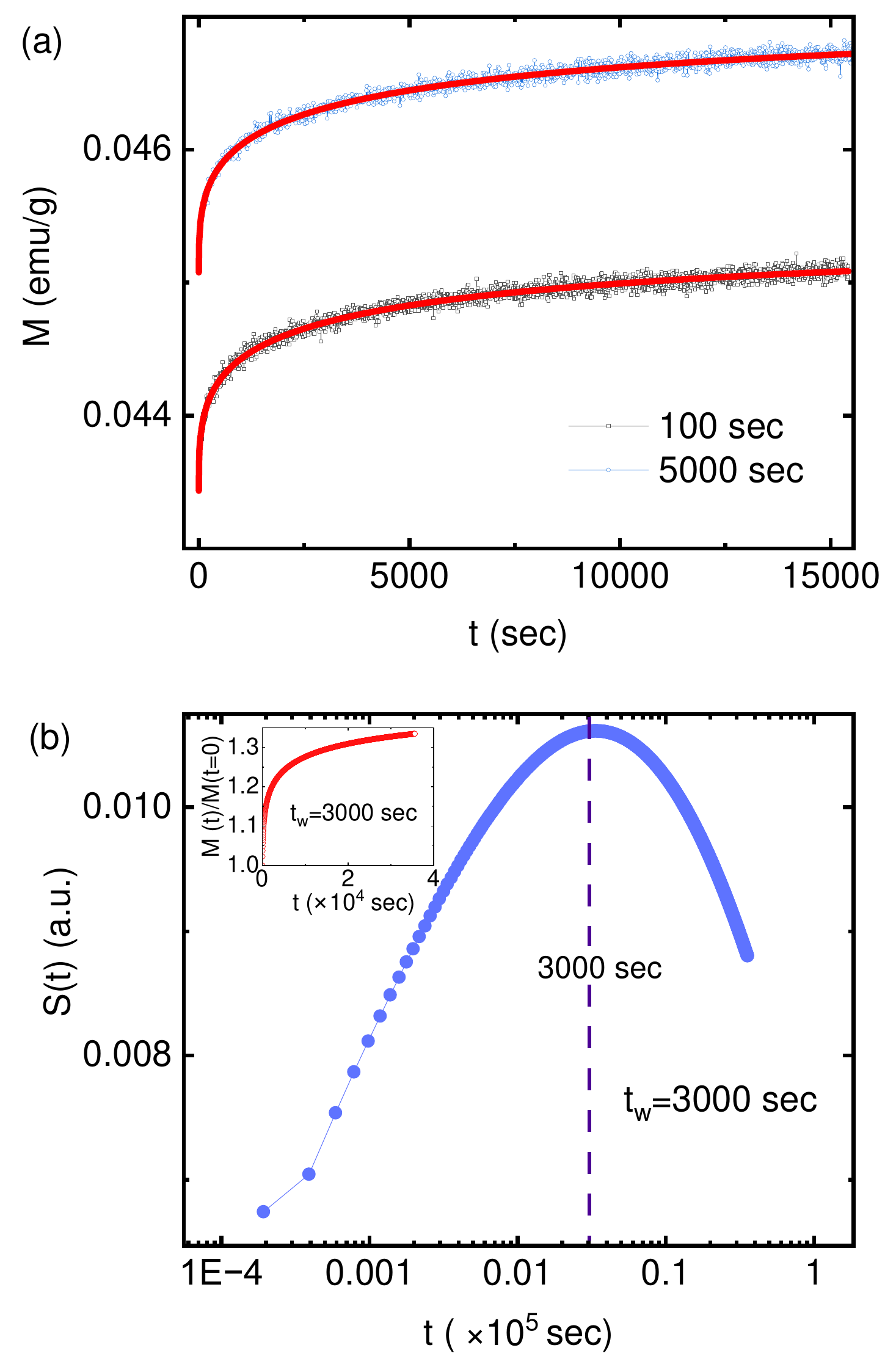}
   \caption{(a) Magnetic relaxation M(t) at 5 K with an applied field of 50 Oe recorded in ZFC procedure for two waiting times t$_w$ = 100  and 5000 s. (b) The relaxation rate S(t) (as described in the text) with the inset showing ZFC magnetic relaxation at 5 K with an applied field of 50 Oe for t$_w$ = 3000 s.}
    \label{Figure 2}
\end{figure}


\subsubsection{\textbf{Magnetic relaxation and aging effect}}
To further characterize the glassy state, time-dependent zero-field-cooled magnetization relaxation M(t) curves were measured. For each relaxation curve, the sample is cooled without applying any field from a high temperature 100 K to a temperature lower than T$_{f}$, i.e., 5 K at 1 K/min and kept at 5 K for a waiting time t$_w$. After waiting for t$_w$, a dc field of 50 Oe is applied and magnetization is recorded as a function of time. The zero field cooled magnetization relaxation M(t) follows the stretched exponential function given below~\cite{PhysRevB.98.144436}:

\begin{equation}
    M(t)=M_0-M_ge^{-(\frac{t}{\tau})^\beta}
    \label{eq 4}
\end{equation}
where M$_0$ is the intrinsic magnetization, M$_g$ is the glassy component of magnetization, $\tau$ is the characteristic relaxation time and $\beta$ is the stretching parameter. For a spin-glass system with a distribution of energy barriers, $\beta$ lies between 0 to 1 where $\beta$=0 indicates that there is no relaxation and $\beta$=1 indicates that the system relaxes with a single time constant. The M(t) curves (shown in FIG.~\ref{Figure 2}(a)) were fitted with the above equation ~\ref{eq 4}. The fitting yields M$_0$=0.045 and 0.047 emu/g, M$_g$=0.002 and 0.002 emu/g, $\tau$=2797 and 3411 s, $\beta$=0.35 and 0.354 for t$_W$=100 and 5000 s respectively. The stretching parameter $\beta$ values match well with those reported for glassy systems~\cite{chu1994dynamic,pakhira2016large}. In spin glass systems, due to the aging phenomena, the magnetic relaxation depends on the waiting time (t$_W$). The aging effect can be conveniently revealed from the relaxation rate S(t)=$dM_{ZFC}/d(lnt)$, obtained from the logarithmic time derivative of the ZFC magnetization. A characteristic feature of the S(t) curve is the inflection point at the observation time close to t$_W$. To investigate the presence of aging, a ZFC magnetization relaxation is performed at 5 K by following the procedure discussed above for t$_W$=3000 s (inset of FIG.~\ref{Figure 2}(b)) and the corresponding S(t) curve is derived. As shown in FIG.~\ref{Figure 2}(b), the S(t) curve exhibits a maximum at 3000 s for the ZFC magnetic relaxation performed after a waiting time of 3000 s. This denotes that the system is in non-equillibrium state, which continuously ages below the freezing temperature and retains a memory of the waiting time.

\begin{figure}
    \centering
    \includegraphics[scale=0.38]{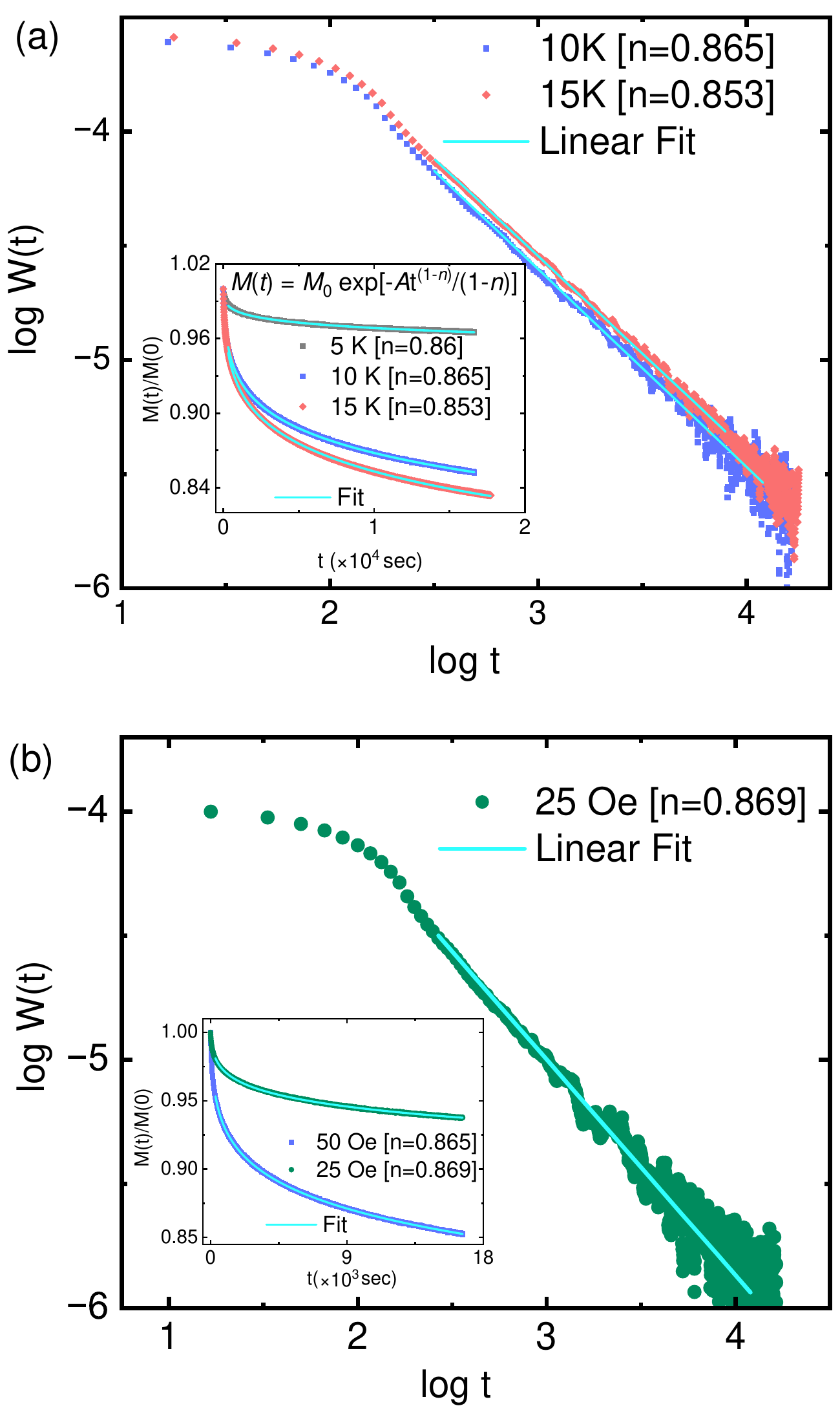}
   \caption{(a) The relaxation rate W(t) at 10  and 15 K with an applied field of 50 Oe in log-log scale obtained from the thermoremanent magnetization relaxation as shown in the inset. (b) The relaxation rate W(t) at 10 K for a different field of 25 Oe in log-log scale obtained from the thermo-remnant magnetization relaxation as shown in the inset. The straight lines represent the linear fits.}
    \label{Figure 3}
\end{figure}

\subsubsection{\textbf{Temperature and field-dependent magnetic relaxation}}

Ulrich \textit{et al.} has proposed a theoretical model based on magnetic particle density dependent relaxation, revealing the glassy phase~\cite{ulrich2003slow}. According to this model, the relaxation rate of thermoremanent magnetization $W(t)=-dlnM(t)/dt$, decays well below the freezing temperature for an assembly of interacting magnetic particles after the lapse of crossover time t$_0$, according to the power law given below

\begin{equation}
W(t)=At^{-n}, t\ge t_0  
    \label{eq 5}
\end{equation}

where A is the prefactor and n is a function of particle density, giving the measure of strength of dipolar interaction. It should be noted that the model by Ulrich \textit{et al.} is valid for the relaxation rate of decay of thermoremanent magnetization, not the ZFC magnetization. The thermoremanent magnetization is measured by cooling the sample from 100 K to the desired temperature (T$_m$=5K, 10K, 15 K) along with applying a field of 50 Oe. After waiting for 3 hours, the field is turned off and magnetization is measured as a function of time as shown in the inset of FIG. ~\ref{Figure 3}(a). The magnetization data measured at 5, 10 and 15 K are fitted with the equation $M(t)=M_0exp[-At^{(1-n)}/(1-n)]$ (since $-dlnM(t)/dt=At^{-n}$) and the value of n is obtained to be 0.86, 0.865 and 0.853 respectively. Further, the relaxation rate W(t) ($=-dlnM(t)/dt$) in the log-log scale  at T$_m$=10 and 15 K is linear fitted as shown in FIG. ~\ref{Figure 3}(a), which also yields $n$=0.86 and 0.85 respectively. Further, the thermoremanent magnetization relaxation is measured at a different field of 25 Oe at 10 K (inset of FIG. ~\ref{Figure 3}(b)) and the value of $n$ is found to be 0.87 from the corresponding fits to magnetization and relaxation rate data (FIG. ~\ref{Figure 3}(b)). The analysis implies that $n$ is almost independent of the measuring temperature and applied field. This is a signature of a canonical spin glass, in contrast to cluster spin glass , where $n$ changes with changing the temperature and field~\cite{rivadulla2004origin}.

\subsubsection{\textbf{Magnetic memory effect}}

\begin{figure}
    \centering
    \includegraphics[scale=0.365]{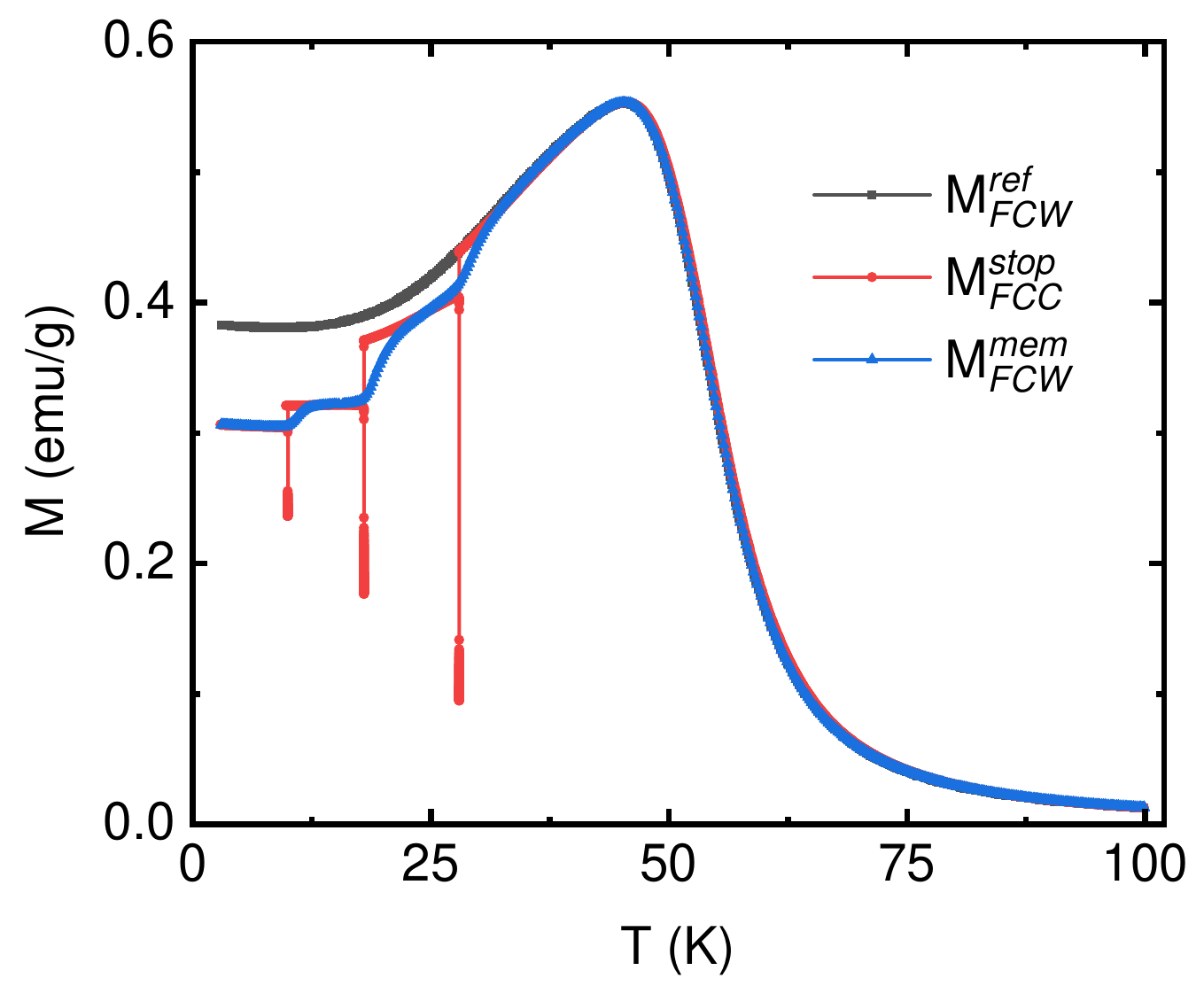}
   \caption{Memory effect in the temperature-dependent field-cooled magnetization at an applied field of 50 Oe. M$_{FCW}^{ref}$ is the reference field-cooled magnetization collected in warming mode. M$_{FCC}^{stop}$ is measured during cooling while stopping at temperatures 10, 18, 28 K for 1.5 h without field, after which the field is turned on, and cooling continues. M$_{FCW}^{mem}$ is measured in the warming cycle after M$_{FCC}^{stop}$ data collection.}
    \label{Figure 4}
\end{figure}

The memory effect is investigated from the field-cooled magnetization curves as shown in FIG. ~\ref{Figure 4}. First, the sample is cooled from 100 K to 3 K at 1 K/min rate by applying a field of 50 Oe. Magnetization is recorded in the warming mode, known as M$_{FCW}^{ref}$. Subsequently, M$_{FCC}^{stop}$ is recorded in the cooling cycle by following the steps mentioned below. The system is cooled from 100  to 3 K at a rate of 1 K/min under an applied field of 50 Oe, with intermediate halting  at 10, 18, and 28 K. At each halting temperature, the field is switched off and the system is allowed to relax for 1.5 h.  Magnetization is recorded continuously throughout the cooling process including these halts. It is found that M$_{FCC}^{stop}$ shows a step-like nature at each halting temperature. where the system was allowed to relax in zero field. After reaching 3 K, the memory curve M$_{FCW}^{mem}$ is collected in the warming mode. M$_{FCW}^{mem}$ shows clear upturns at the halting temperatures 10, 18, 28 K, trying to mimic M$_{FCC}^{stop}$ curve. This clearly indicates that M$_{FCW}^{mem}$ remembers the previous magnetic history~\cite{khan2014memory,pakhira2016large} and provides strong evidence of a prominent memory effect in this system.

\begin{figure}
    \centering
    \includegraphics[scale=0.37]{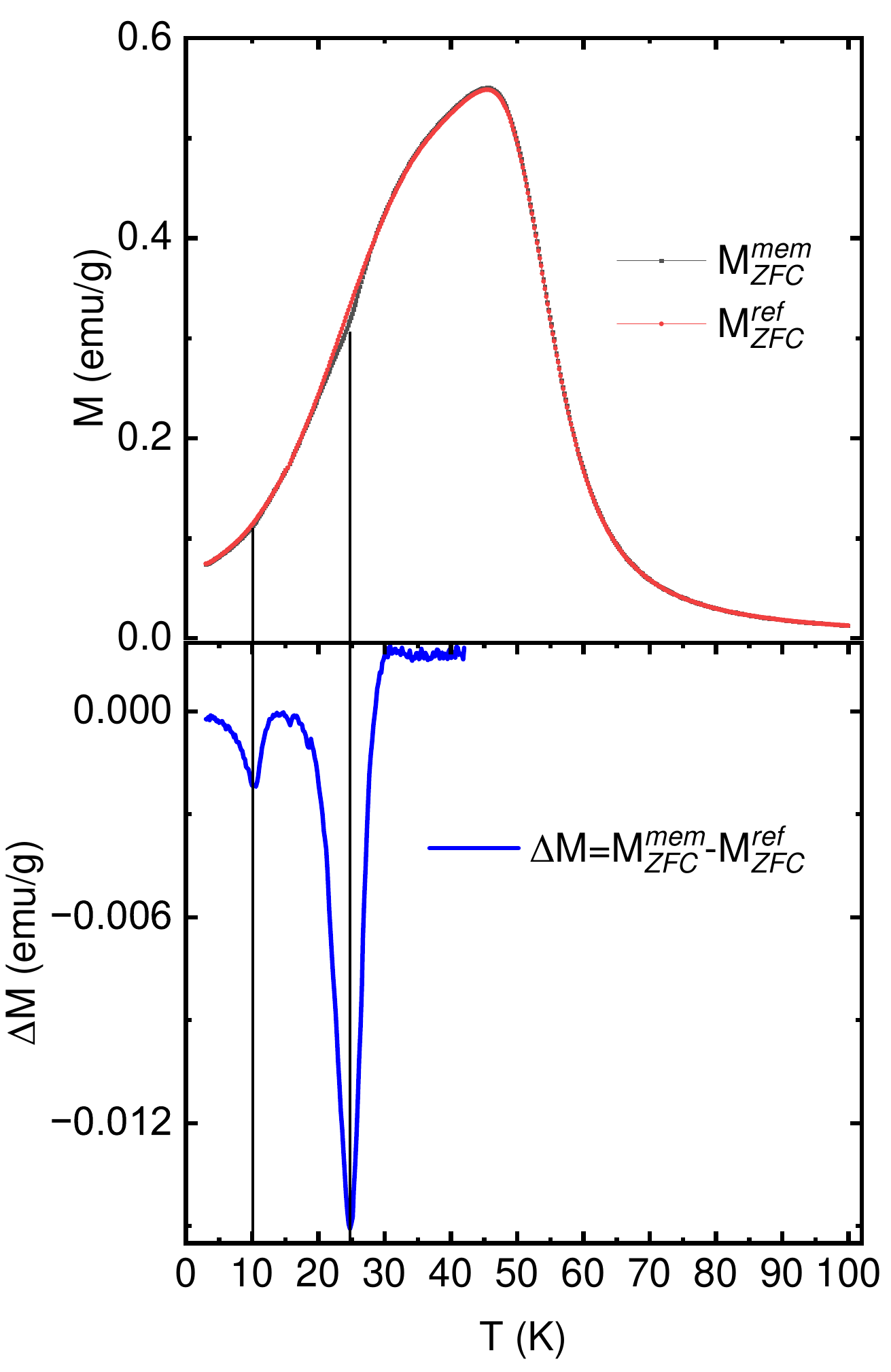}
   \caption{Upper panel: Temperature dependent zero field cooled magnetization (M$_{ZFC}^{ref}$) and zero field cooled magnetization with memory effect (M$_{ZFC}^{mem}$). Lower panel: Magnitude difference ($\Delta$M) between M$_{ZFC}^{mem}$ and M$_{ZFC}^{ref}$.}
    \label{Figure 5}
\end{figure}

It is worth noting that both spin-glass and superparamagnetic systems can exhibit memory effects in field-cooled magnetization, whereas memory effects in zero-field-cooled magnetization are usually stronger signature of spin-glass systems~\cite{sasaki2005aging}. Therefore, the memory effect is further substantiated in the zero-field-cooled magnetization measurements as discussed below. First a conventional zero-field-cooled curve is recorded with an applied field of 50 Oe (FIG. ~\ref{Figure 5}). Then, to investigate the memory effect, the sample is rapidly cooled from 100 K to 3K at zero field while halting at temperatures 10 K and 25 K for 2.5 hours. Then the magnetization (M$_{ZFC}^{mem}$) is recorded in the warming cycle after applying a field of 50 Oe. M$_{ZFC}^{mem}$ shows a characteristic memory dip at the halting temperatures 10  and 25 K. The dip is clearly visible at 10  and 25 K in the $\Delta$M (M$_{ZFC}^{mem}$-M$_{ZFC}^{ref}$) curve. The presence of memory dips in the zero-field-cooled curve further corroborates the presence of glassy behavior in this compound.

\begin{figure}
    \centering
    \includegraphics[scale=0.38]{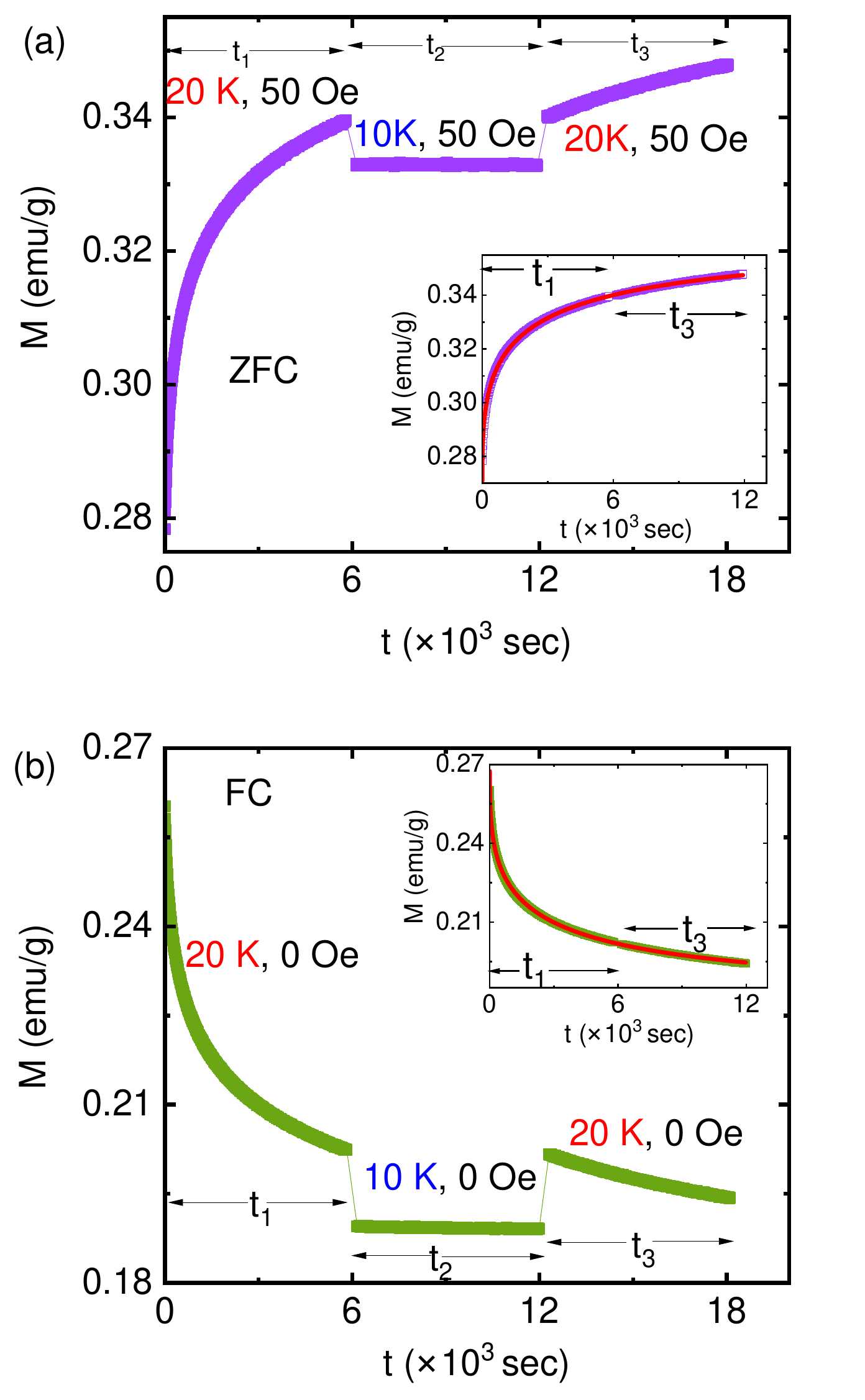}
   \caption{Magnetic relaxation at 20 K with intermediate cooling at 10 K for (a) ZFC(b) FC procedure with 50 Oe applied field. The inset shows that the relaxation during t$_3$ is a continuation of the relaxation during t$_1$, with the red curve representing the fit with equation ~\ref{eq 4}.}
    \label{Figure 6}
\end{figure}

The mechanism of memory effect in glassy systems can be described via two models; the droplet model~\cite{fisher1988nonequilibrium,yoshino2001multiple} and hierarchical model~\cite{dotsenko1985fractal,lefloch1992can}. While the droplet model includes a unique equilibrium single spin configuration at a certain temperature, the hierarchical model involves a complex free energy landscape comprising multiple metastable states~\cite{khan2014memory,pakhira2016large}. Experimental signature of these two models lies with respect to the recovery of the original spin configuration after a positive temperature cycle. In the droplet model, the original spin configuration is restored when the system is cooled back to its initial temperature after a temporary heating, thereby retaining memory of the initial magnetic aging. By contrast, in the hierarchical model, temporary heating rejuvenates the system, resulting in the loss of memory effect. In the hierarchical model, each valley in the free-energy landscape gets further divided into sub-valleys when temperature is lowered, whereas these sub-valleys progressively merge with increase in temperature. When the temperature decreases from T to T-$\Delta$T, each valley splits into multiple subvalleys. If $\Delta$T is large, then the energy barrier between the sub-valleys becomes high. Hence, the system can not overcome these barriers and the relaxation occurs within the subvalleys of each valley. When the temperature goes back to T, the subvalleys and barriers merge to the original multivalley structure on the free energy landscape. Hence, the relaxation at T is not affected by the intermediate cooling at T-$\Delta$T. In contrast, when temperature increases from T to T+$\Delta$T, the barriers between the valleys gets lowered and even the valleys gets merged. This allows the system to escape from its initially occupied valleys and access other metastable states, resulting in relaxation over a different free energy landscape. However, when the temperature is brought back to the original temperature T after the intermediate heating, though the free energy landscape is restored, the relative occupancy in each valley is different and the original spin configuration can not be restored. Hence, the hierarchical model accounts for the asymmetric magnetic relaxation for positive and negative temperature cycle.

\begin{figure}
    \centering
    \includegraphics[scale=0.36]{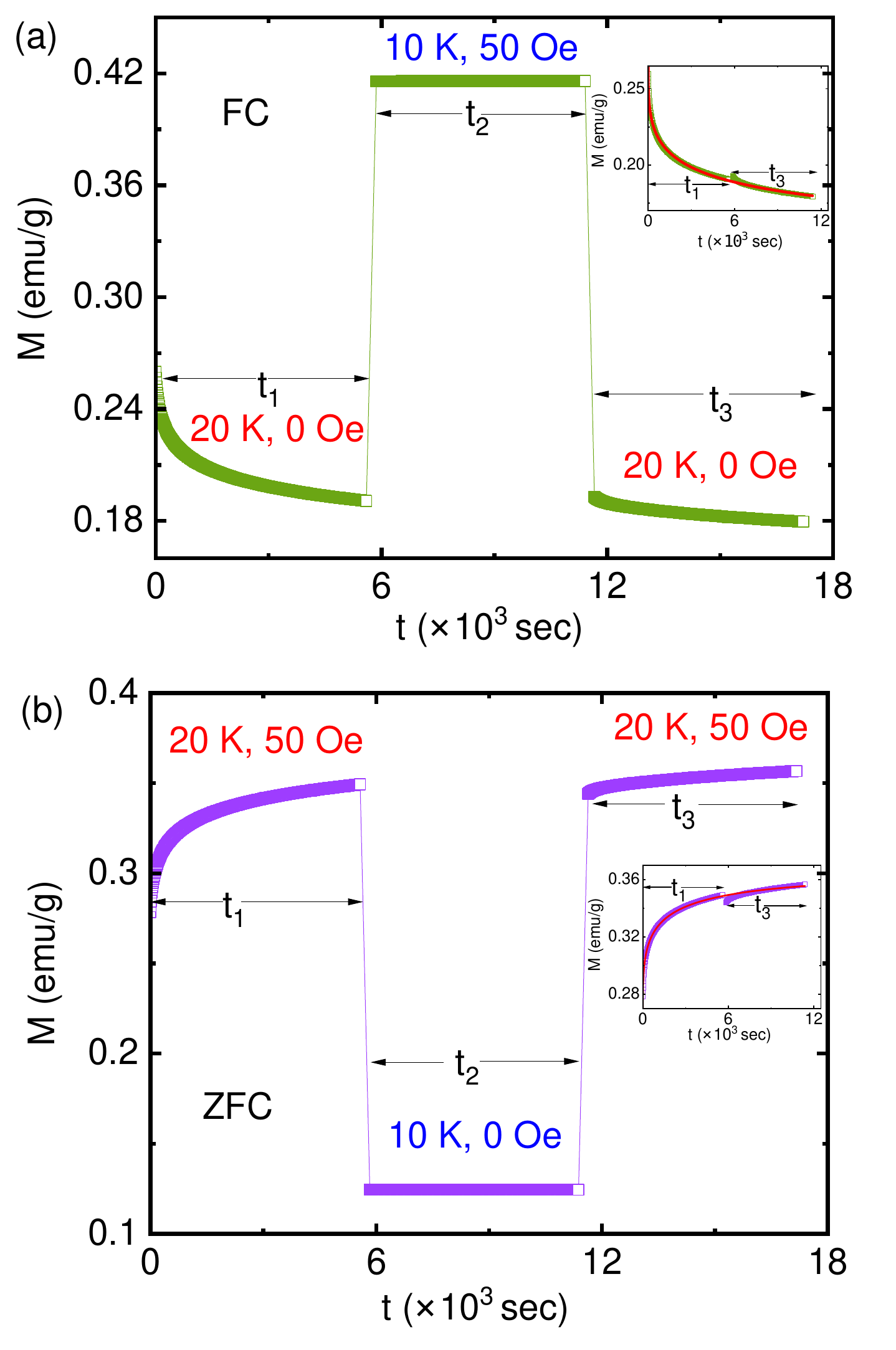}
   \caption{Magnetic relaxation at 20 K with intermediate cooling at 10 K and change in applied field for (a) FC(b) ZFC procedure with 50 Oe applied field. The inset shows that the relaxation during t$_3$ is a continuation of the relaxation during t$_1$, with the red curve representing the fit with equation ~\ref{eq 4}.}
    \label{Figure 7}
\end{figure}

Taking this into consideration, we have investigated the effect of temperature cycling on magnetization relaxation by following the protocol proposed by Sun \textit{et al.}~\cite{sun2003memory}. First, the ZFC and FC magnetization relaxation measurements with a negative temperature cycle (intermediate cooling) are performed by utilizing two different protocols as shown in FIG. ~\ref{Figure 6}(a)-(b) and FIG. ~\ref{Figure 7}(a)-(b). In the first protocol, the applied field remains constant throughout the relaxation process (FIG. ~\ref{Figure 6}(a)-(b)), whereas, in the second protocol, the applied field is changed with intermediate cooling (FIG. ~\ref{Figure 7}(a)-(b)).

\begin{figure}
    \centering
    \includegraphics[scale=0.35]{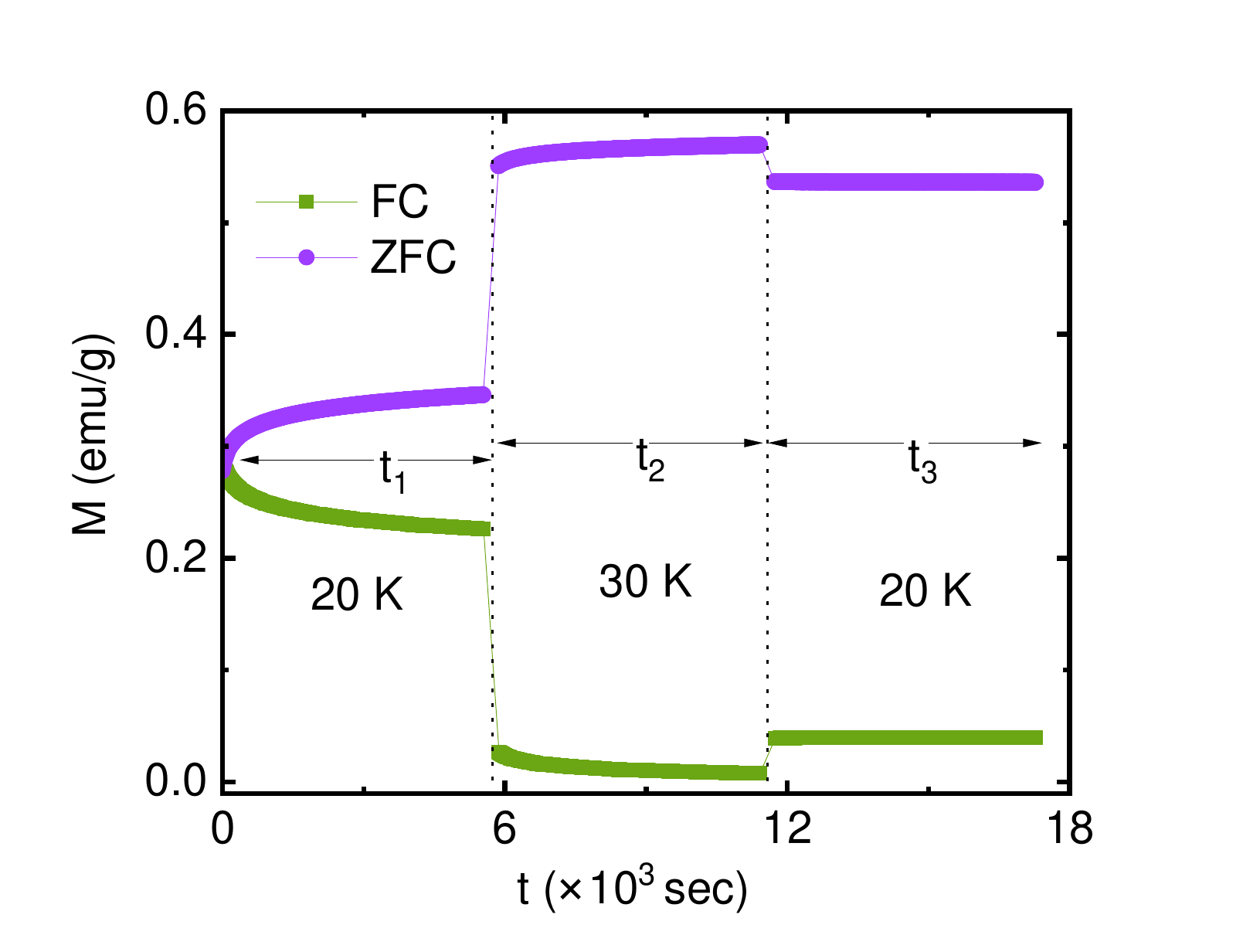}
   \caption{Magnetic relaxation at 20 K with intermediate heating at 30 K (positive temperature cycle) at 50 Oe applied field for FC and ZFC methods.}
    \label{Figure 8}
\end{figure}

Following the first protocol, in the ZFC mode (FIG. ~\ref{Figure 6}(a)), the sample is cooled from 100 K to 20 K at 1K/min without applying any field. After reaching 20 K, 50 Oe field is applied and magnetization is recorded as a function of time for t$_1$ sec. After that, the sample is cooled to 10 K without changing the field and magnetization is recorded for t$_2$ sec. Then the sample is again heated back to 20 K and magnetization is recorded for t$_3$ sec. Throughout the entire relaxation process, the applied field is kept constant. In the FC method (FIG. ~\ref{Figure 6}(b)), the sample is cooled from 100 K to 20 K at 1K/min with applying 50 Oe field. After reaching 20 K, the field is turned off and similar intermediate cooling is performed while recording magnetization as a function of time. It should be noted that, after the intermediate cooling, both the FC and ZFC magnetizations attain the nearly same value, which it had just before the cooling. The insets show that the relaxation during t$_3$ is a continuation of relaxation during t$_1$ and the whole curve can be fitted with a single equation using the stretched exponential function (equation ~\ref{eq 4}), suggesting the presence of a strong memory effect.

To further verify the memory effect using the second protocol mentioned above, we have performed relaxation measurements for both FC (FIG. ~\ref{Figure 7}(a)) and ZFC method (FIG. ~\ref{Figure 7}(b)) by turning the field on and off respectively during the time interval t$_2$. As shown in FIG. ~\ref{Figure 7}, with intermediate cooling during t$_2$, the field is changed from 0 to 50 Oe in FC and 50 to 0 Oe in ZFC method. However, despite this field switching during intermediate cooling, still the magnetization relaxation during t$_3$ attains the almost same initial value, which it had just before the cooling during t$_2$ and the relaxation during t$_1$ and t$_3$ can be fitted with the same stretched exponential equation ~\ref{eq 4}. This indicates that the memory effect in this material is quite robust below the freezing temperature.

To demonstrate the effect of a positive temperature cycle, magnetic relaxations are recorded with intermediate heating in this compound for both ZFC and FC method by maintaining the applied field constant throughout the relaxation process (FIG. ~\ref{Figure 8}). In the ZFC method, the sample is cooled from 100 K to 20 K at 1K/min without applying any field. After reaching 20 K, 50 Oe field is applied and magnetization is recorded as a function of time for t$_1$ sec. After that, the sample is heated to 30 K without changing the field and magnetization is recorded for t$_2$ sec. Then the sample is finally cooled back to 20 K and magnetization is recorded for t$_3$ sec. In the FC mode, the sample is cooled from 100 K to 20 K at 1K/min with applying 50 Oe field. After reaching 20 K, the field is turned off and similar intermediate heating is performed while recording magnetization as a function of time. Here, it should be noted that, the magnetization during relaxation at t$_3$ has not attained the value, which it had just before the heating during t$_2$ and the relaxation curves during t$_1$ and  t$_3$ look quite different. Hence the intermediate heating rejuvenates the magnetic relaxation and there is no memory effect in this compound with intermediate heating protocol. The asymmetric response between positive and negative temperature cycle suggests the presence of hierarchical organization of metastable states and this compound favors hierarchical model over the droplet model.

\section{CONCLUSION}
We report the detailed structural and magnetic characterization in (Fe$_{0.55}$Ni$_{0.45}$)$_3$GaTe$_2$. The underlying magnetic interaction in this compound is found to be ferromagnetic. The irreversibility observed in dc magnetization along with the frequency dependent peak shift in ac susceptibility suggests the presence of a spin glass state in this system. Further, the relative shift in freezing temperature along with the characteristic parameters obtained using power law and Vogel-Fulcher law  determine the canonical spin glass nature of the system. The non-equillibrium dynamics studies via magnetic relaxation and memory effect measurements using different experimental techniques show the presence of glassy behavior, aging phenomena and a strong memory effect. The presence of memory effect in ZFC magnetization reveals that the memory effect arises from a spin-glass rather than superparamagnetism. The asymmetric magnetic relaxation response with respect to positive and negative temperature cycle favors the hierarchical model of memory effect over the droplet model.

\textit{Acknowledgments}-P.B. and D.S. acknowledge financial support from the DAE APEX Project, IoP, Bhubaneswar. PM acknowledges the support from the Raja Ramanna Chair (RRC) scheme of the Department of Atomic Energy, India.

\nocite{*}

\bibliography{apssamp}

\end{document}